\documentclass[12pt]{iopart}

\usepackage{graphicx}
\usepackage[all]{xy}
\usepackage{epstopdf}
\begin{document}

\title{Mathematical Methods in Quantum Optics: the Dicke Model \\}

\author{Eduardo
Nahmad-Achar, Octavio Casta\~nos, Ram\'on L\'opez-Pe\~na, and Jorge G. Hirsch}

\address{Instituto de Ciencias Nucleares, Universidad Nacional
Aut\'onoma de M\'exico, Apdo. Postal 70-543 M\'exico 04510 D.F.}

\ead{nahmad@nucleares.unam.mx}

%\vspace{0.2in}

\begin{abstract}

\noindent We show how various mathematical formalisms, specifically the catastrophe formalism and group theory, aid in the study of relevant systems in quantum optics.

We describe the phase transition of the Dicke model for a {\it finite} number $N$ of atoms, via $3$ different methods, which lead to universal parametric curves for the expectation value of the first quadrature of the electromagnetic field and the expectation value of the number operator, as functions of the atomic relative population. These are valid for {\it all} values of the matter-field coupling parameter, {\it and} valid for both the ground and first-excited states.

Using these mathematical tools, the critical value of the atom-field coupling parameter is found as a function of the number of atoms, from which its critical exponent is derived.

\end{abstract}

\pacs{42.50.Ct, 03.65.Fd, 64.70.Tg}

\maketitle

%%%%%%%%%%%%%%%%%%%%%%%%%%%%%%%%%%%%%%%%%%%%
%% MAINMATTER
%%%%%%%%%%%%%%%%%%%%%%%%%%%%%%%%%%%%%%%%%%%%

\section{Introduction}

With the ability to manipulate single atoms and photons in a cavity came a renewed interest in the models that describe their behaviour. The Hamiltonian for matter and field interactions can be written in two equivalent forms~\cite{power}, viz., the one associated to the multipolar expansion 
	\begin{equation}
	H_M = \sum_\alpha \left( \frac{\vec{\pi}_\alpha^2}{2 m} 
	+ V(r_\alpha) \right)  -e \sum_\alpha  
	\left( \vec{E}\cdot \vec{r}_\alpha \right) +
	\frac{1}{8 \pi} \int{ \bigl(\vec{E}^2 + \vec{B}^2 \bigr) \, d^3 \vec{r}}
	\end{equation}
where the first term describes the atomic field, the second the dipole interaction with the electric field, and the third the radiation field itself; and the one related to the radiation gauge
	\begin{equation}
	H_R = \frac{\vec{p}_\alpha^2}{2 m}  + V(r_\alpha) -\frac{e}{m \, c} 
	\sum_\alpha \left( \vec{p}_\alpha \cdot \vec{A}\right) + 
	\frac{e^2}{2 \, m \, c^2} \vec{A}^2 +
	\frac{1}{8 \pi} \int{ \bigl(\vec{E}^2 + \vec{B}^2 \bigr) \, d^3 \vec{r}} \ \  
	\end{equation}
where the long wavelength approximation is considered, and which contains a so-called {\it diamagnetic term} quadratic in the electromagnetic vector potential $A$. 	They are both related via the (unitary) gauge transformation~\cite{power,siva}
	\begin{equation}
		U = \exp[i\frac{e}{\hbar c} \sum_{s=1}^{N} r_s \cdot A]\, ,
	\end{equation}
i.e., both Hamiltonians give rise to the same physical predictions when treated correctly.

When approximating to $2$-level systems, the multipole Hamiltonian gives rise to the well known Dicke model (DM)~\cite{dicke}. This describes the interaction of a single mode quantized radiation field with a sample of $N$ two-level atoms, located inside an optical cavity, in the dipolar approximation (i.e., located within a distance smaller than the wavelength of the radiation), and has the form
	\begin{equation}
	\label{fullhamiltonian}
H_D = \frac{1}{N}\hbar \omega_F a^\dagger a + \frac{1}{N} \tilde{\omega}_A J_z + \frac{\tilde{\gamma}}{\sqrt{N}N} (a^\dagger J_{-} + a J_{+}) + \frac{\tilde{\gamma}}{\sqrt{N}N} (a^\dagger J_{+} + a J_{-})\ \ 
	\end{equation}
where $\omega_F$ is the field frequency, $\hbar\tilde{\omega}_A$ is the atomic energy separation, $a^{\dagger}\ \mbox{and}\ a$ denote the one-mode creation and annihilation photon operators respectively, $J_{z}$ the atomic relative population operator, and $J_{\pm}$ the atomic transition operators. We have taken $\hbar = 1$. The interaction Hamiltonian (third and fourth terms on the right hand side) contains, respectively, the {\it rotating} and {\it counter-rotating} terms, with coupling constant $\tilde{\gamma}$. Note that we have divided the Hamiltonian by the total number of particles $N$, having in this way an intensive Hamiltonian operator. This is convenient for the thermodynamic limit is found by just taking $N\rightarrow\infty$, though our interest is to study the behaviour at {\it finite} $N$.

The approximation to $2$-level systems of the radiation Hamiltonian $H_R$, however, yields a {\it different} Hamiltonian $H_{R\,2\ell}$ which contains the diamagnetic term and which is {\it not} gauge-equivalent to $H_D$. This is summarised in the following non-commutative diagram
\begin{equation*}
	\xymatrixcolsep{5pc}
	\xymatrix@R+2em@C+2em{
	H_M \ar[r]^{2-level\ approx} \ar[d]_-U & H_D \ar@{-->}[d]^{\exists \!\!\!/} \\
	H_R \ar[r]_{2-level\ approx} & H_{R\,2\ell}
 }
\end{equation*}
where the downward dashed arrow signifies that no gauge transformation exists between the two.

We stress here the distinction between $H_D$ and $H_{R\,2\ell}$ because there has been confusion in the literature. An important feature of the Dicke Hamiltonian $H_D$ is the presence of a phase transition from the normal to a {\it super-radiant} behaviour~\cite{hepp}. This is a collective effect involving all $N$ atoms in the sample, where the decay rate is proportional to $N^2$ instead of $N$, the expected result for independent atom emission. This phenomenon was observed experimentally in optically pumped HF gas~\cite{exp1}. However, the transition from the normal to the super-radiant regime has been much debated in the literature, mainly because when using $H_{R\,2\ell}$ the Thomas-Reiche-Kuhn sum rule would place contradictory bounds to the parameters of the model (cf. e.g.~\cite{wod1}), and because it was believed that gauge invariance {\it requires} the presence of the diamagnetic term~\cite{knight}. (The essence behind these arguments lies in the fact that the coupling strength $\tilde{\gamma}$ is much smaller than the atomic level separation $\tilde{\omega}_A$, for optical systems.) At variance with this common belief, it has more recently been argued that the Dicke model does give a consistent description of the interaction of a one-mode light field with the internal excitation of atoms inside a cavity~\cite{domokos}. Furthermore, recent experimental results indicate that it can actually be observed~\cite{baumann,nagy} by using Raman transitions between ground states in an atomic ensemble, thus circumventing the obstacle $\tilde{\gamma}\ll\tilde{\omega}_A$.

In general, if we write $H_D = H_0 + V$, with $U_0 = \exp[-(i/\hbar)\,H_0\, t]$ the evolution operator for $H_0$, then, since $U_0^\dagger a U_0 = a\, \exp[-i(\omega_F/N)\,t]$ and $U_0^\dagger J_+ U_0 = J_+\, \exp[-i(\tilde{\omega}_A/N)\,t]$, we have
\begin{eqnarray}
	V_I &=& U_0^\dagger\, V\, U_0 \nonumber \\
	&=& \frac{\tilde\gamma}{N^{3/2}} \left( a^\dagger e^{i \frac{\omega_F}{N} t} J_- e^{-i \frac{\tilde{\omega}_A}{N} t} + a\, e^{-i \frac{\omega_F}{N} t} J_+ e^{i \frac{\tilde{\omega}_A}{N} t} \right) + \nonumber \\ 
	&\ & + \frac{\tilde\gamma}{N^{3/2}} \left( a^\dagger e^{i \frac{\omega_F}{N} t} J_+ e^{i \frac{\tilde{\omega}_A}{N} t} + a\, e^{-i \frac{\omega_F}{N} t} J_- e^{-i \frac{\tilde{\omega}_A}{N} t} \right) \nonumber \\
	&=& \frac{\tilde\gamma}{N^{3/2}} \left( a^\dagger J_- e^{\frac{i}{N}(\omega_F-\tilde{\omega}_A) t} + a\,J_+ e^{-\frac{i}{N}(\omega_F-\tilde{\omega}_A) t} \right) + \nonumber \\
	&\ & + \frac{\tilde\gamma}{N^{3/2}} \left( a^\dagger J_+ e^{\frac{i}{N}(\omega_F+\tilde{\omega}_A) t} + a\,J_- e^{-\frac{i}{N}(\omega_F+\tilde{\omega}_A) t} \right)
\end{eqnarray}

When $\omega_F$ and $\tilde{\omega}_A$ are close to one another (as when in or close to resonance), and of the order of optical frequencies ($\sim 10^{15}s^{-1}$), the terms of the form $e^{\pm i(\omega_F + \tilde{\omega}_A)t}$ vary rapidly in a time-scale of $1/\omega_F$ with an average close to zero, and may be neglected. This is the well-studied Tavis-Cummings~\cite{tavis} model (TCM) (cf. e.g. \cite{scrip, scrip2, vaxjo2010}), and the approximation is called the {\it rotating wave} approximation. 

\begin{figure}[h]
\begin{center}
\scalebox{0.75}{\includegraphics{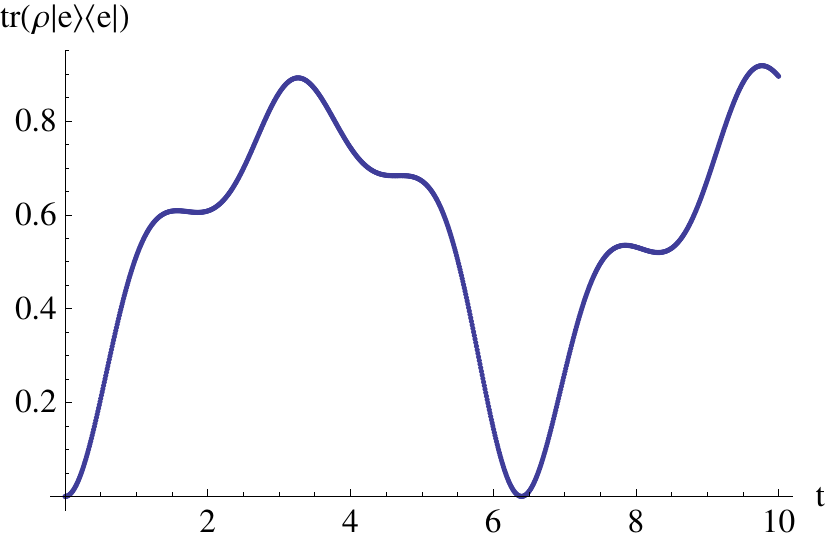}} \quad
\scalebox{0.75}{\includegraphics{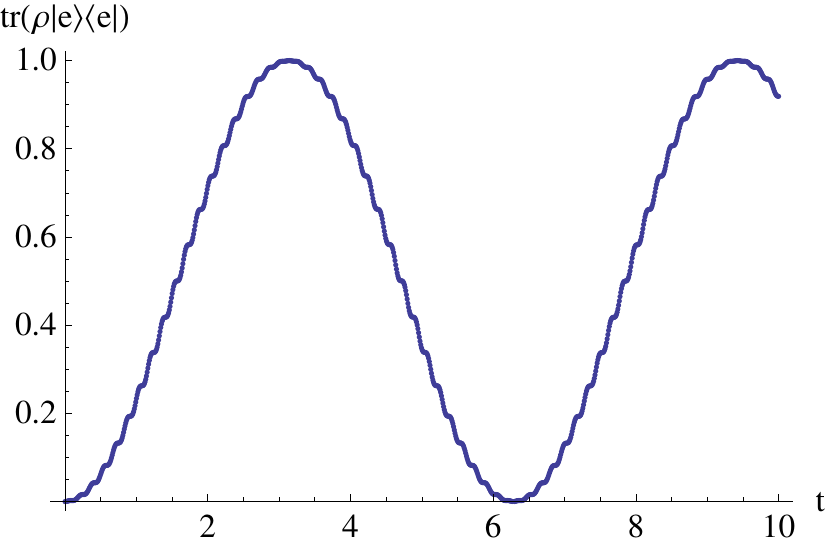}}
\label{evol}
\caption{Evolution of an excited atom. The plots show the probability of finding the atom in its excited state, as a function of time, without using the rotating wave approximation. When $\tilde{\omega}_A$ is small (left, $\tilde{\omega}_A = \Omega$) high frequency terms deviate the result from the harmonic behaviour obtained analytically in the rotating wave approximation. These terms tend to disappear when $\tilde{\omega}_A$ is large (right, $\tilde{\omega}_A = 19\,\Omega$).}
\end{center}
\end{figure}

Figure~\ref{evol} shows this very clearly~\cite{canela}, where the time evolution of the excited state $\vert e\rangle$ of one atom has been plotted for $\tilde{\omega}_A = \omega_F = \Omega$ (left), and $\tilde{\omega}_A = \omega_F = 19\,\Omega$ (right), with $\Omega = \tilde\gamma / (N^{3/2}\hbar)$ the Rabi frequency of the atom; here, the time scale has been taken in units of $1/\Omega$. As $\tilde{\omega}_A$ increases, the evolution approaches the harmonic oscillations that can be obtained analytically in this approximation.

%Figure 1: evolution of excited state in RWA

The TCM model has more recently been generalised~\cite{dukelsky}, and considered to describe cavity QED with Bose-Einstein condensates~\cite{brennecke}. The simplest case $N=1$ in the rotating-wave approximation is known as the Jaynes-Cummings model (JCM)~\cite{jc1963}. Many theoretical predictions of this latter model, such as the existence of collapse and revivals in the Rabi oscillations~\cite{eberly}, the formation of macroscopic quantum states, or measures of entanglement associated with spin-squeezed states, have been confirmed, and many experimental studies of Rydberg atoms with very large principal quantum number within single-mode cavities have been realised~\cite{manko}. Together, the JCM and the TCM have served as a guide to understand several quantum optics phenomena.

In this work we will consider the full Dicke Model, with rotating and counter-rotating terms, and for simplicity we define $\omega_A = \frac{\tilde\omega_A}{\omega_F},\ \gamma = \frac{\tilde\gamma}{\omega_F},\ j = \frac{N}{2}$. We shall also measure frequencies in units of the field frequency, i.e. $\omega_F=1$, so that
	\begin{equation}
		\label{hamiltonian}
			H = \frac{1}{N} a^\dagger a + \frac{\omega_A}{N} J_z + 		\frac{\gamma}{N \sqrt{N}}(a^\dagger + a) (J_{+} + J_{-})
	\end{equation}
and $\omega_A,\, \gamma,\ J_{+},\, J_{-},\, J_z,\, J_x,\, J_y$ are all adimensional. The coupling parameter $\gamma$ is given, in terms of physical quantities of the system, as $\gamma=\sqrt{(2\pi\varrho/\hbar \,\tilde{\omega}_{F})}\,\vec{d}_{ba}\cdot\vec{e}_{P}$, where $\varrho$ denotes the density of atoms in the volume of quantisation, $\vec{d}_{ba}$ is the excitation matrix element of the electric dipole operator of a single atom, and $\vec{e}_{P}$ the polarisation vector.

We will show how to calculate the phase transition of the Dicke model for a {\it finite} number $N$ of atoms, via $3$ different methods: i) numerical diagonalisation of the Hamiltonian and the use of the fidelity between neighbouring states; ii) variational test states that are a direct product of coherent Heisenberg-Weyl ($HW(1)$)-states (for the electromagnetic field), and $SU(2)$-states (for the atomic field)~\cite{papercorto,paperextenso}; and iii) the use of projection operators on the coherent states in (ii) to obtain states which obey the parity symmetry in the total excitation number, present in the Hamiltonian. It will be seen that the {\it symmetry-adapted} states (iii) constitute a much better approximation (than the simple coherent states) to the exact quantum ground and first-excited states, apart from offering the localisation of a {\it precursor} of the quantum phase transition for {\it finite} $N$ (usually referred to in the literature as a quantum phase transition in a finite system~\cite{leviatan}).

It is rather surprising that the three different approaches mentioned above lead to universal parametric curves for the expectation values of both the first quadrature of the electromagnetic field, and the atomic relative population, as functions of the atom-field coupling parameter. These are valid for the ground- and first-excited states, and are also presented.

Finally, using these procedures, the critical value of the atom-field coupling parameter $\gamma_{c}$ may be found as a function of the number of atoms $N$, from which its critical exponent may be derived.

\section{Coherent States}

The systems described by Eq.(\ref{hamiltonian}) are non-integrable and must be solved via numerical diagonalisation for specific scenarios, but good semiclassical approximations that mimic well the exact quantum behaviour have been constructed~\cite{papercorto,paperextenso,emary}.

One of these is to take as a test state a direct product of coherent $HW(1)$-states $\vert\alpha\rangle$ (for the electromagnetic field), and $SU(2)$-states $\vert\zeta\rangle$ (for the atomic field):
   \begin{equation}
	      \vert\alpha\rangle
	      \otimes\vert\zeta\rangle=\frac{e^{
	      -\left|\alpha\right|^{2}/2}}{
	      \left(1+\left|\zeta\right|^{2}\right)^{j}}
	      \sum_{\nu=0}^{\infty}\sum_{m=-j}^{+j} \left\{
	      \frac{\alpha^{\nu}}{\sqrt{\nu!}}\,
	      {2j \choose j+m}^{1/2}\,
	      \zeta^{j+m}\,\,\vert\nu\rangle\otimes\vert j,\,m\rangle
	      \right\}
   \end{equation}
here $\nu$ represents the eigenvalues of the photon number operator $a^\dagger a$, $m$ those of the atomic relative population operator $J_{z}$, and $j(j+1)$ those of the squared total angular momentum operator $J^2$.
We can calculate the expectation values of the field and matter observables by using the representation of the angular momentum and Weyl generators with respect to the tensorial product $\vert\alpha\rangle\otimes\vert\zeta\rangle$~\cite{hecht}. Acting with the operator $J_z$ on an arbitrary state $\vert\psi\rangle$ and projecting it onto the $SU(2)$ coherent state $\vert\zeta\rangle$, we obtain
\begin{eqnarray}
\langle\zeta\vert J_z\vert\psi\rangle &=& \frac{1}{(1+\vert\zeta\vert^2)^j}\,\langle j,\,-j\vert\, e^{\zeta^* J_-} J_z\, \vert\psi\rangle \nonumber \\
&=& \frac{1}{(1+\vert\zeta\vert^2)^j} \left( -j + \zeta^* \frac{\partial}{\partial\zeta^*} \right) \{\zeta\vert\psi\rangle
\end{eqnarray}
where we have used $\frac{\partial}{\partial\zeta^*}\, e^{\zeta^* J_-} = J_- \,e^{\zeta^* J_-}$, and written $\vert\zeta\} = (1+\vert\zeta\vert^2)^j \,\vert\zeta\rangle$ for the unnormalised state. The second term in parenthesis may be calculated as the operator
$$
\frac{1}{(1+\vert\zeta\vert^2)^j}\, \zeta^* \frac{\partial}{\partial\zeta^*} = \frac{j \zeta\zeta^*}{(1+\vert\zeta\vert^2)^{j+1}} + \zeta^* \frac{\partial}{\partial\zeta^*}\left[ \frac{1}{(1+\vert\zeta\vert^2)^j}\,\, \cdot \right]
$$
so that, substituting above, we get
$$
\langle\zeta\vert J_z\vert\psi\rangle = \left( -\frac{j}{1+\vert\zeta\vert^2} + \zeta^* \frac{\partial}{\partial\zeta^*} \right) \langle\zeta\vert\psi\rangle
$$
i.e., the representation for $J_z$ in the basis of coherent states is
\begin{equation}
J_z \rightarrow -\frac{j}{1+\vert\zeta\vert^2} + \zeta^* \frac{\partial}{\partial\zeta^*}
\end{equation}
Similarly, for the other operators we have
\begin{eqnarray}
J_+ &\rightarrow& j \, \zeta^* \,
\frac{2 + \vert \zeta\vert^2}{1 + \vert \zeta\vert^2} +
(\zeta^*)^2 \, \frac{\partial}{\partial \zeta^*} \nonumber \\
J_- &\rightarrow& \frac{j \, \zeta}{1 + \vert
        \zeta\vert^{2}}+\frac{\partial}{\partial \zeta^*} \nonumber\\
a &\rightarrow& \frac{\partial}{\partial \alpha^*} +\frac{\alpha}{2} \nonumber\\
a^\dagger &\rightarrow& \alpha^*
\end{eqnarray}

The {\it energy surface} is then easily found by taking the expectation value of the Dicke Hamiltonian with respect to $\vert\alpha\rangle\otimes\vert\zeta\rangle$, and the best variational approximation to the ground state energy of the system, as well as its corresponding eigenstate, are obtained from it. In our case the energy surface has the form	
   \begin{eqnarray}
      {\cal H}(\alpha,\,\zeta)&\equiv&
      \langle\alpha\vert\otimes\langle\zeta\vert\,\,
      H\,\,\vert\alpha\rangle\otimes\vert\zeta\rangle\nonumber\\
      &=&\frac{1}{2N}\left(p^{2}+q^{2}\right)-\frac{\omega_A}{2}\cos\theta +
 	  \frac{\gamma}{\sqrt{j}} q\,\sin\theta\,\cos\phi
   \end{eqnarray}
where we have defined $\alpha=\frac{1}{\sqrt{2}}\left(q+i\,p\right),\ 
      \zeta=\tan\left(\frac{\theta}{2}\right)
      \,\exp\left(i\,\phi\right),$ and where $(q,\,p)$ correspond to the expectation values of the radiation field quadratures, and $(\theta,\,\phi)$ determines a point on the Bloch sphere.

\begin{figure}[h]
\begin{center}
\scalebox{0.75}{\includegraphics{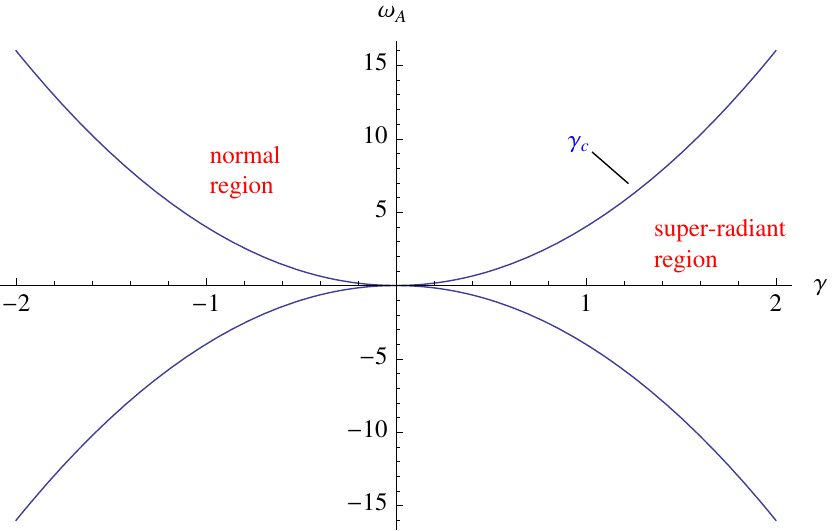}}
\label{separatrix}
\caption{Separatrix for the normal and super-radiant regions given by the critical points obtained from the coherent states.}
\end{center}
\end{figure}

Critical points which minimise the energy surface are obtained via the catastrophe formalism~\cite{gilmore}
	 \begin{eqnarray*}
		\theta_{c}=0,\ q_{c}=0,\ p_{c}=0 \hfill
		&&\hbox{for}\ \vert\gamma\vert<\gamma_{c} \nonumber \\
		\theta_{c}=\arccos(\gamma_c/\gamma)^{2},\ 
		q_{c}=-2\,\sqrt{j}\,\gamma\,
		\sqrt{1-(\gamma_c/\gamma)^{4}}\cos{\phi_c},\  
		p_{c}=0&&,\ \phi_{c}=0,\,\pi \nonumber \\
		&&\hbox{for}\ \vert\gamma\vert>\gamma_{c} \nonumber
	\end{eqnarray*}
where $\gamma_c=\sqrt{\omega_a/4}$. This defines the {\it separatrix} of the phase transition, as shown in Fig.~\ref{separatrix}. Note that, for the CS used, it is independent of $N(=2j)$.

%Figure 2: separatrix

\begin{figure}
		\scalebox{0.8}{\includegraphics{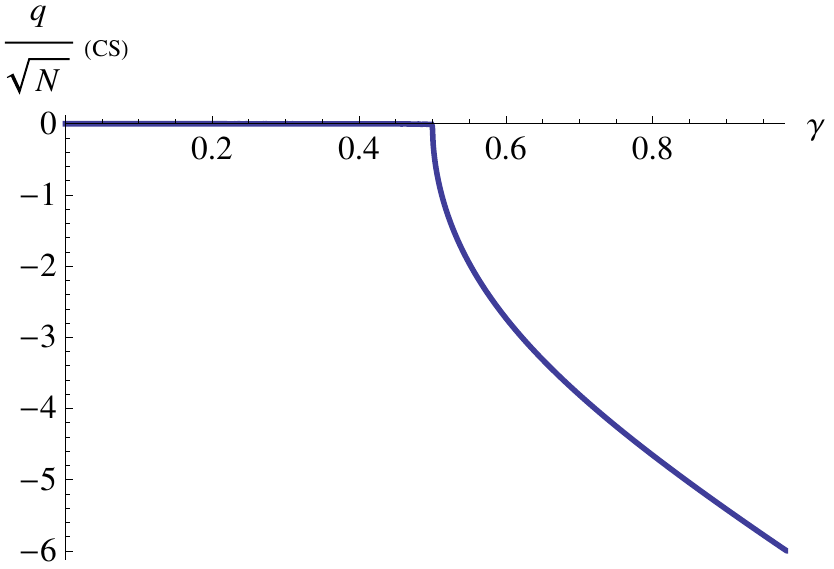}} \quad
		\scalebox{0.8}{\includegraphics{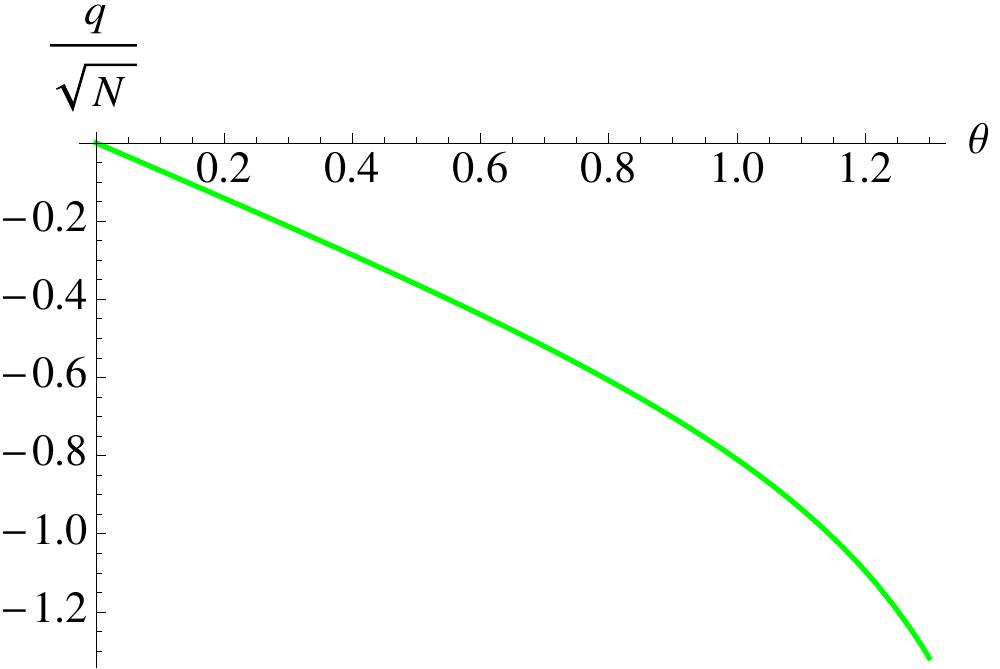}}
		\caption{Left: Critical value of $\frac{q_c}{\sqrt{N}}$ as a function of the interaction strength $\gamma$. Right: $q_c/\sqrt{N}$ versus $\theta_c$ for {\it all} $\gamma$ and {\it any} number of atoms. In both curves $\omega_{A}=1,\ \phi_c=0$.}
		\label{qgammatheta_cs}
	\end{figure}

Figure~\ref{qgammatheta_cs} (left) shows the critical value for the (normalised) field quadrature $q_c/\sqrt{N}$ as a function of the interaction strength $\gamma$. The presence of a phase transition from the normal (where this quantity vanishes) to the super-radiant phase is evident. From the expressions for the critical points we obtain, in the super-radiant region,
	\begin{equation}
				\frac{q_c}{\sqrt{N}}= - \sqrt{\omega_A} 		\frac{\sin\theta_c}{\sqrt{2\,\cos\theta_c}}\cos\phi_c
	\end{equation}
Figure~\ref{qgammatheta_cs} (right) shows this relationship; the normal regime is described only by the origin $(q_c=0,\, \theta_c=0)$. The rest of the curve belongs to the super-radiant regime. Note that this curve is valid for {\it all} $\vert\gamma\vert > \gamma_c$ (as this parameter drops out from the equation above). It is also valid for {\it any} number $N$ of atoms. We shall see in what follows that this curve is universal.

%Figure 3: q, \theta vs.\gamma, && universal curve for CS
	
A natural variable to define is $x=\gamma/\gamma_c$, and the energy at the minima take the values
		\begin{equation}
	    	E_{\hbox{normal}}=-2N\,\gamma_{c}^{2}\ ,\quad
	            E_{\hbox{super-radiant}}=-N\,\gamma_{c}^{2}\,x^{2}\,\left(
	            1+x^{-4}\right)
	            \label{ecoherente}
		\end{equation}
One may study the statistical properties of the variational states by calculating the expectation values of linear matter and field observables with respect to the CS, at the minima, as well as their fluctuations~\cite{paperextenso}. For instance, for the eigenvalues $\lambda$ of the excitation number operator $\hat{\Lambda}=\hat{a}^{\dagger}\hat{a}+\hat{J}_{z}
+\sqrt{\hat{J}^{2}+1/4}\,-1/2$, we have
		\begin{equation}
	    	\lambda_{\hbox{normal}}=0\, ,
	            \quad\lambda_{\hbox{super-radiant}}=\frac{N}{2}\,\left[1
	            -x^{-2}+2\,\gamma_{c}^{2}\,x^{2}\,\left(1-x^{-4}\right)
	            \right]
	            \label{lambda}
		\end{equation}
The fluctuations $\Delta\hat\Lambda$ are zero in the normal region,
and in the super-radiant phase take the form
		\begin{equation}
		        \Delta\hat{\Lambda}=\sqrt{\frac{N}{2}
			\left(\frac{1}{2}+2\,\gamma^{2}_c \, x^2 \right)\left(1-x^{-4}
			\right)}
		\end{equation}
		
A complete table for the expectation values and fluctuations of the most important matter and field observables is given in \cite{paperextenso}. Many of these resemble very well the results that one would obtain by solving numerically the Hamiltonian eigenvalue equation for the ground and first-excited states (more on this later). The real beauty of it all is that analytic expressions can be obtained for all these values. But some of them, however, fail to approximate the exact quantum results. Notably, the fluctuations of the quadrature operator $\Delta\hat{q}$ and of the atomic transition operator $\Delta\hat{J}_x$, obtained from the CS, show an erroneous behaviour. The reason can be traced down to the fact that, while the Hamiltonian (\ref{hamiltonian}) shows a parity symmetry given by
	\begin{equation}
		[e^{i \pi \hat{\Lambda}},\, H] = 0
		\label{symmetry}
	\end{equation}
the coherent states mix states with opposite symmetry.

\section{Symmetry-Adapted States}

Considering the unitary transformations $U(\chi) = \exp\left(i\,\chi\,\hat{\Lambda}\right)$, and using the relations $U(\chi)\,a\,U^{\dagger}(\chi) = e^{-i\,\chi}\,a$, $U(\chi)\,J_{+}\,U^{\dagger}(\chi) = e^{-i\,\chi}\,J_{+}$, we find
\begin{eqnarray}
U(\chi)\,H\,U^{\dagger}(\chi) = && a^{\dagger}a
+\omega_{A}J_{z}+\frac{\gamma}{\sqrt{N}}
\left(a^{\dagger}\,J_{-}+a\,J_{+}\right) \nonumber \\
&& +\frac{\gamma}{\sqrt{N}}
\left(e^{-2i\chi}\,a^{\dagger}\,J_{+}+
e^{2i\chi}\,a\,J_{-}\right)
\end{eqnarray}
so we have a {\it symmetry transformation} for $\chi = 0,\,\pi$. I.e., $H$ is invariant under the group ${\cal C}_{2}=\left\{I, e^{i\,\pi\hat{\Lambda}}\right\}$.

This symmetry allows for the classification of the Hamiltonian eigenstates in terms of the parity of the eigenvalues $\lambda = \nu + m + j$ of $\Lambda$, whereas states with opposite symmetry are strongly mixed by the variational CS states considered in the last section. We may, however, obtain {\it symmetry-adapted states} (SAS) ~\cite{papercorto,paperextenso} by projecting with $P_{\pm}=\frac{1}{2}\left(I \pm e^{i\,\pi\,\hat{\Lambda}}\right)$, i.e.,
	\begin{eqnarray}
			\vert\alpha,\,\zeta\rangle_{\pm}&=&
			{\cal N}_{\pm}\,\,P_{\pm}\,
			\vert\alpha,\,\zeta\rangle \nonumber \\
			&=&{\cal N}_{\pm}\,\left[ \ 
			\vert\alpha\rangle\otimes\vert\zeta\rangle\pm\
			\vert-\alpha\rangle\otimes\vert-\zeta\rangle\ \right]\
	\end{eqnarray}
with ${\cal N}_{\pm}$ the normalisation factors for the {\it even} $(+)$ and {\it odd} $(-)$ states. The {\it surface energy} takes a more complicated form
		\begin{eqnarray}
			&&\langle H\rangle_\pm=\pm\frac{1}{2}\left(p^2+q^2\right)
			\left\{1-\frac{2}{1\pm e^{\pm(p^2+q^2)}(\cos\theta)^{\mp
			N}}\right\} \nonumber \\
			&&\quad-\frac{N}{2}\,\omega_{A}
	 		\left\{(\cos \theta)^{\pm 1} 
	 		\pm\frac{\tan^2\theta\,\cos\theta}{1\pm e^{\pm(p^2+q^2)} 
			(\cos\theta)^{\mp N} }\right\} \nonumber \\
			&&\quad+\sqrt{2\,N}\,\gamma\left\{\frac{\pm p\,\tan\theta\, 
			\sin\phi+q\,e^{p^2+q^2}
			\sin\theta\,\cos\phi\,(\cos\theta)^{-N}}{
			e^{p^2+q^2}(\cos\theta)^{-N}\pm 1}\right\}
			\label{symad}
	\end{eqnarray}
and the critical points which minimise it can be obtained, this time numerically, following the same method as for the coherent states (CS). In this case $p_{c}=0;\ \phi_{c}=0,\,\pi;$ and $q_c$ and $\theta_c$ now depend strongly on $N$. [Note: one could still obtain analytic expressions if, as usually done in many-body physics, one were to use the critical points for the CS energy surface, in order to obtain the trial state which approximates the two lowest symmetry-adapted energy states, and in which to evaluate the expectation values of the observables. 	Around the phase transition there would be observable differences, however, and one should calculate the critical points of (\ref{symad}) as we do here. For the first approach see ~\cite{paperextenso}\,].

\begin{figure}
	\scalebox{0.87}{\includegraphics{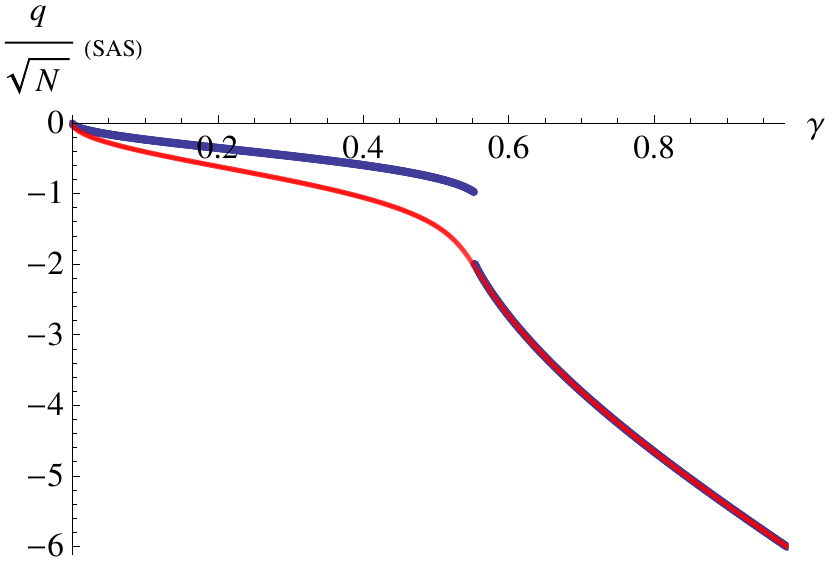}} \quad
	\scalebox{0.8}{\includegraphics{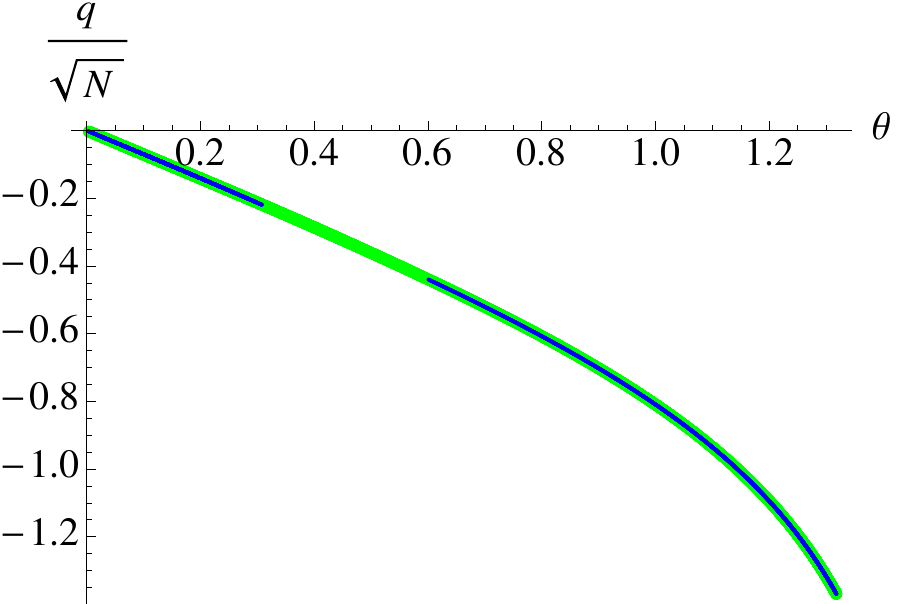}}
	\caption{(Colour online.) Left: $\frac{q_c}{\sqrt{N}}$ as function of the interaction strength $\gamma$, for the ground (blue, discontinuous upper curve) and first-excited (red, continuous lower curve) SAS states. Here, $N=20$, $\omega_{A}=1$, and $\phi_c=0$. Right: $q_c/\sqrt{N}$ as a function of $\theta_c$ for the CS and odd SAS states (continuous, green) and even SAS state (discontinuous, blue).}
	\label{q_vs_gamma_SAS}
	\end{figure}

\begin{figure}
    \scalebox{0.75}{\includegraphics{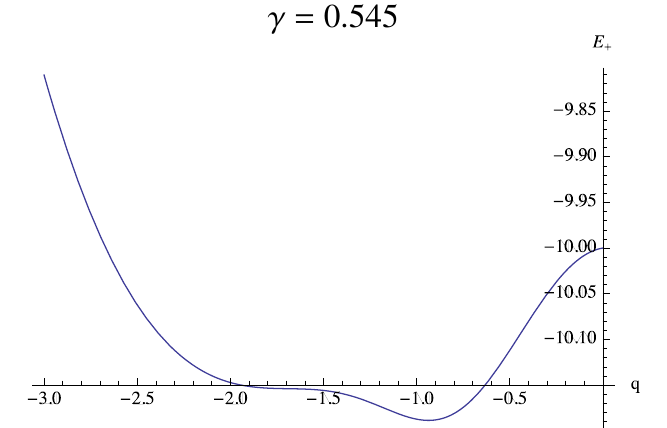}} \quad
    \scalebox{0.75}{\includegraphics{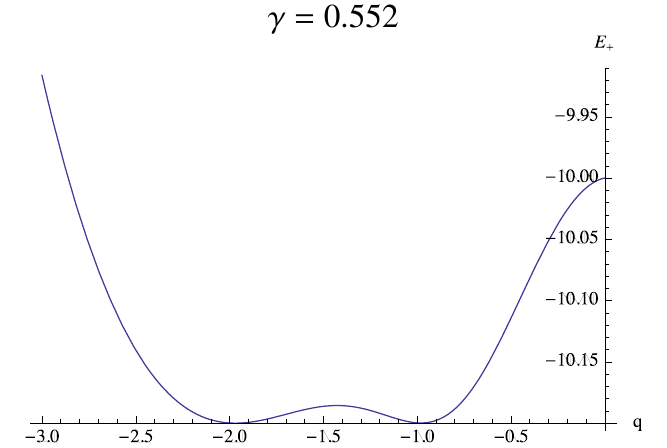}} \quad
	\scalebox{0.75}{\includegraphics{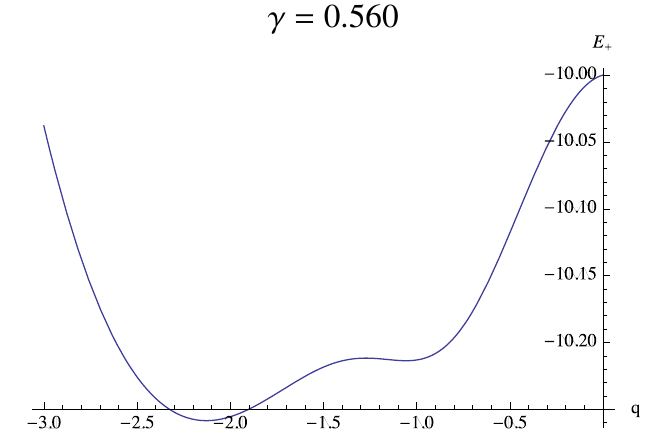}}
\caption{The symmetry adapted energy surface showing the minima for $\gamma= 0.545, 0.552, 0.560$ plotted as a function of $q$. Same values for $N$ and $\omega_A$ as before.}
\label{discontinuity}
\end{figure}

Figure~\ref{q_vs_gamma_SAS} (left) shows the behaviour of $q_c$ vs. $\gamma$ for the ground (even, blue, discontinuous upper curve) and first-excited (odd, red, continuous lower curve) states. The discontinuity in the phase space variable for the even state marks the phase transition~\cite{universal}. This discontinuity arises from a competition for the global minimum in the even energy surface between two local minima, as $\gamma$ crosses the phase transition region, which can be seen in Fig.~\ref{discontinuity}\,~\cite{rumania}. In the case of the odd states, an inflection point marks the transition from one regime to the other.

%Figure 4: q_vs_gamma SAS, and universal CS-SAS

%Figure 5: competition for global minimum SAS

If we compare the minimum points $q_c$ associated to the energy surfaces of the CS and the even SAS, the discontinuity mentioned above translates into a forbidden range of values for $\theta_c$ ($0.3\le\theta\le 0.6$) when $q_c$ is plotted against it in Fig.~\ref{q_vs_gamma_SAS} (right). But it is interesting to note that, in the allowed $\theta$-region, both fall along the same curve. The odd SAS state is also plotted, falls along the same curve, has no forbidden regions, and is indistinguishable from that of the coherent state.

Finally, it must be mentioned that the SAS states have been successfully used to study phase transitions through the R\'enyi-Wehrl entropy~\cite{romera}.

\section{Exact Solution}

\begin{figure}
		\scalebox{1.3}{\includegraphics[width=2.2in]{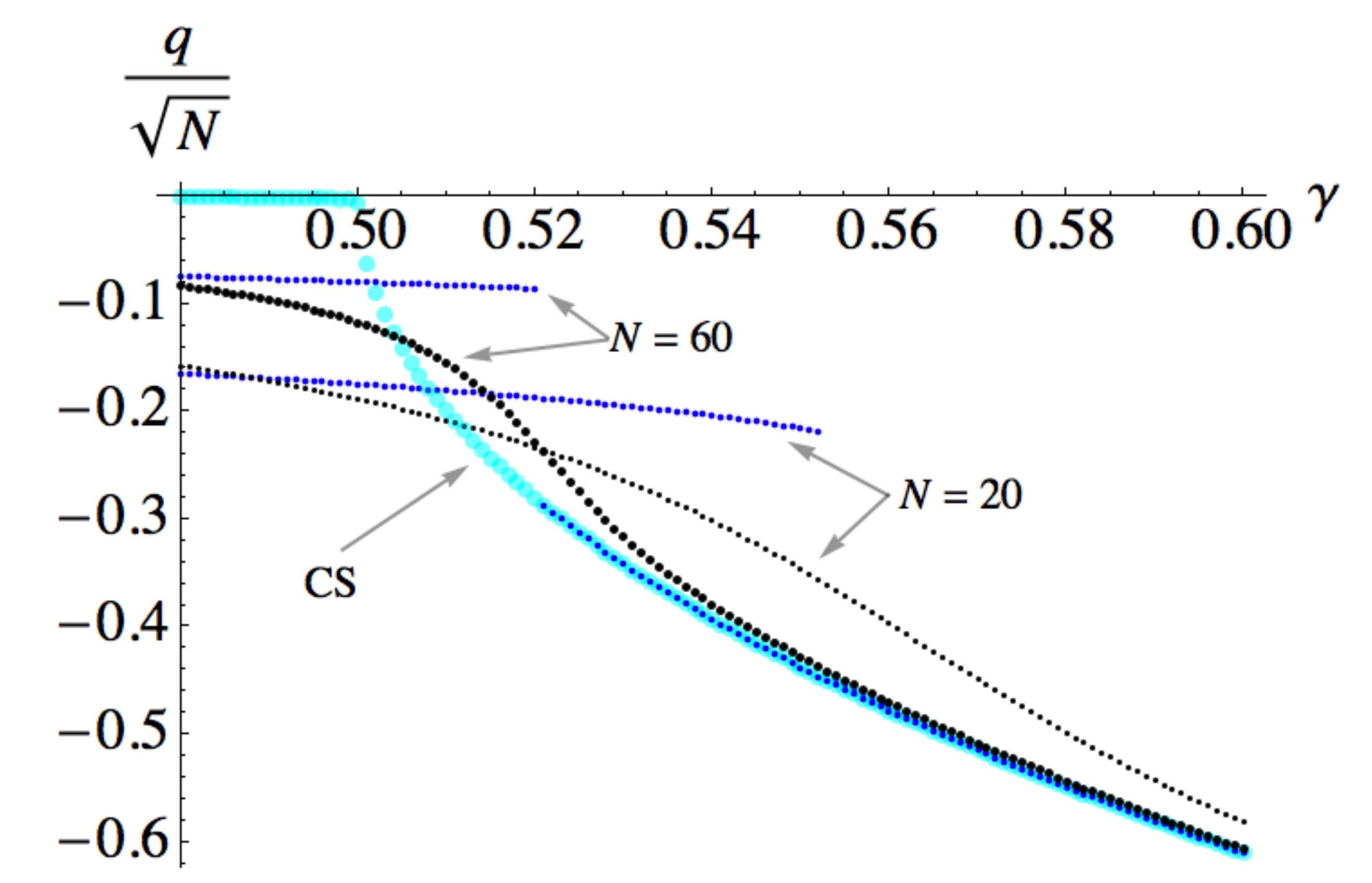}} \qquad
		\scalebox{1.35}{\includegraphics[width=2.2in]{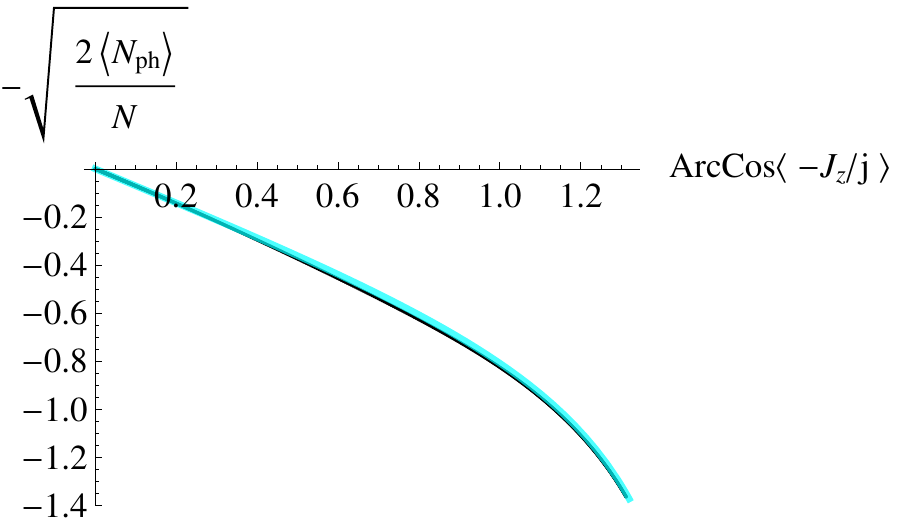}}
		\caption{(Colour online.) Left: dependence with $\gamma$ of the quadrature (expectation value of the number operator) for $N=20$ (bottom continuous gray dotted line), and $N=60$ (top continuous black dotted line), compared with those for the SAS state at the same values of $N$ (discontinuous dotted lines) and with the result for the CS (top continuous cyan line). Right: Universal curve for quantum, SAS, and CS states.}
		\label{Exact_SAS_CS_univ}
	\end{figure}

In order to study exact quantum solutions we propose the correspondence
$q_c  \rightarrow  \mp \sqrt{2 \langle a^{\dagger}a\rangle}$, $\theta_c \rightarrow \arccos\left({ \frac{-\langle J_{z}\rangle}{j}}\right)$, where
$\mp$ corresponds to $\phi_c=0, \, \pi$. As the system is not integrable, we solve numerically the Hamiltonian eigenvalue equation for the ground and first-excited states, separating the even- and odd-parity cases. The quantum phase transition may be found through the fidelity between neighbouring states~\cite{universal}~\cite{zanardi}
$$
\mathcal{F} = 		\vert\langle\psi(\gamma)\,\vert\,\psi(\gamma+\delta\gamma)\rangle|^2\, ,
$$
which at the transition acquires its minimum value. An associated, more precise method, is to use the {\it fidelity susceptibility} (the second derivative of the fidelity function given above) which is maximum at the phase transition region~\cite{gu}. The dependence with $\gamma$ of the quadrature (or expectation value of the number operator) is shown in Figure~\ref{Exact_SAS_CS_univ} (left), for $N=20$ (bottom continuous gray dotted line) and $N=60$ (top continuous black dotted line). These are compared with those for the SAS state at the same values of $N$ (discontinuous dotted lines) and with the result for the CS (top continuous line). Note that the quantum and SAS solutions agree at $\langle N_{ph}\rangle = q_c = 0$, and approximate very well beyond the phase transition, while the CS fails to reproduce results well enough in the normal region. Recall that the SAS curves are discontinuous at the phase transition due to a competition for the global minimum in the even energy surface between two local minima, as explained in the last Section. After the discontinuity, the curves for any value of $N$ become indistinguishable with the CS curve, as the figure shows. As $N$ increases, $\gamma_c \rightarrow 0.5$ and the jump in the SAS curves becomes smaller; these curves approach more and more their quantum counterparts. Also, for greater values of $\gamma$ all curves tend to one and the same curve.

%Figure 6: q vs \gamma and universal curve for all states

One can also plot $q_c$ vs. $\theta_c$ (or $\sqrt{\langle a^{\dagger}a\rangle/j}$ vs. $\arccos(\langle -J_{z}\rangle/j)$) for the quantum state (Fig.~\ref{Exact_SAS_CS_univ} (right)) falling indistiguishably on the same universal curve together with the CS and SAS even and odd states.

\section{Critical Exponents}

A function $f(r)$ is called {\it homogeneous} if, for all values of $\lambda$, we have $f(\lambda r) = g(\lambda)\,f(r)$. The function $g(\lambda)$ is called the {\it scaling function} and must be of the form $g(\lambda) = \lambda^s$; $s$ is called the {\it critical exponent}. The scaling hypothesis states that the singular part of many potentials are homogeneous functions near second-order phase transitions; this is true, for instance, for all thermodynamic potentials~\cite{katzgraber}.

Critical exponents thus serve to describe the behaviour of important observables of a system near phase transitions and, although in most cases their properties are only supported by experimental data, they are believed to be universal with respect to physical systems.

\begin{figure}
	\scalebox{0.75}{\includegraphics{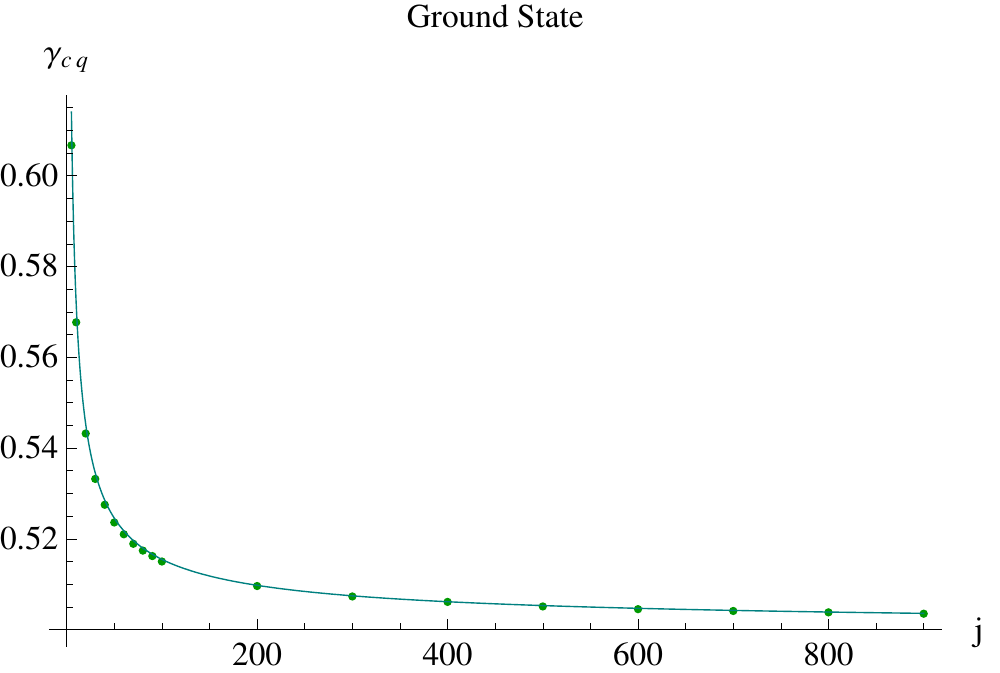}} \quad
	\scalebox{0.75}{\includegraphics{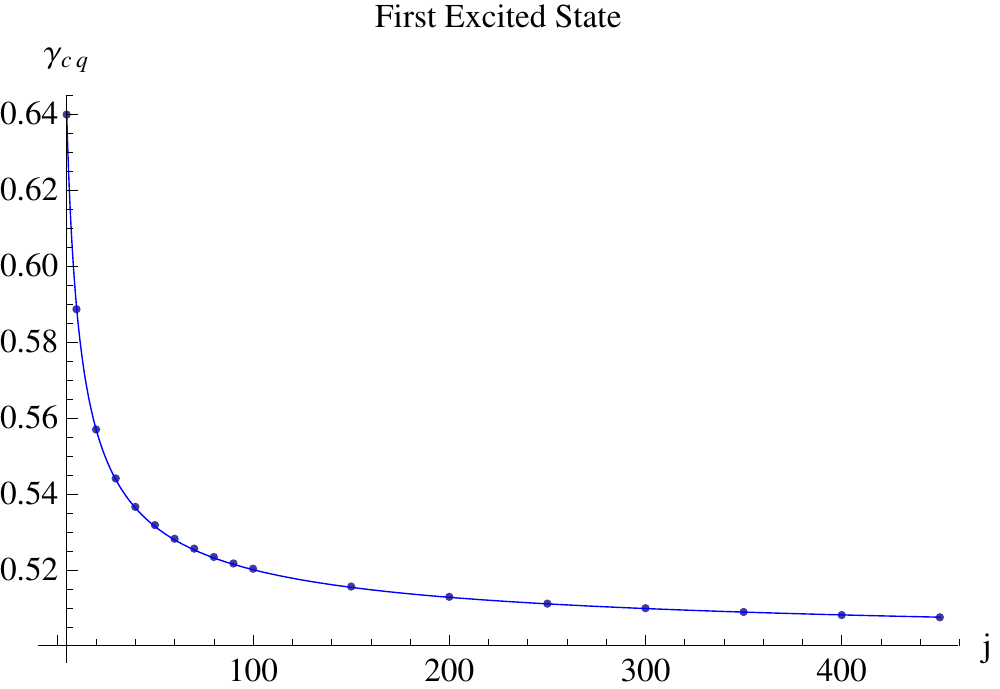}}
	\caption{Behaviour of the critical value of the coupling parameter $\gamma_c$ with the number of atoms $N\,(=2j)$, for the quantum ground (left) and first-excited (right) states. In both cases $\gamma_c$ grows as $N^{-2/3}$.}
	\label{gamma_vs_j}
\end{figure}

Our treatment for {\it finite} $2$-level systems in a cavity, in the presence of a radiation field, allows us to study the critical value of the atom-field coupling parameter $\gamma_{c}$ as a function of the number of atoms $N$, from which its critical exponent may be derived. Figure~\ref{gamma_vs_j} shows this relationship for the exact, quantum ground (left) and first-excited (right) states. The points correspond to the numerical solution obtained from the diagonalisation of the Hamiltonian, while the continuous curves are the fits
\begin{eqnarray}
	\gamma_c = \frac{1}{2} + \frac{1}{3}\, j^{-\frac{2}{3}} \ :
	&& \hbox{(ground state)} \\
	\gamma_c = \frac{1}{2} + \frac{2}{5}\, j^{-\frac{2}{3}} \ :
	&& \hbox{(first-excited state)}
\end{eqnarray}

%Figure 7: q_vs_j quantum

\begin{figure}
	\begin{center}
	\scalebox{0.9}{\includegraphics{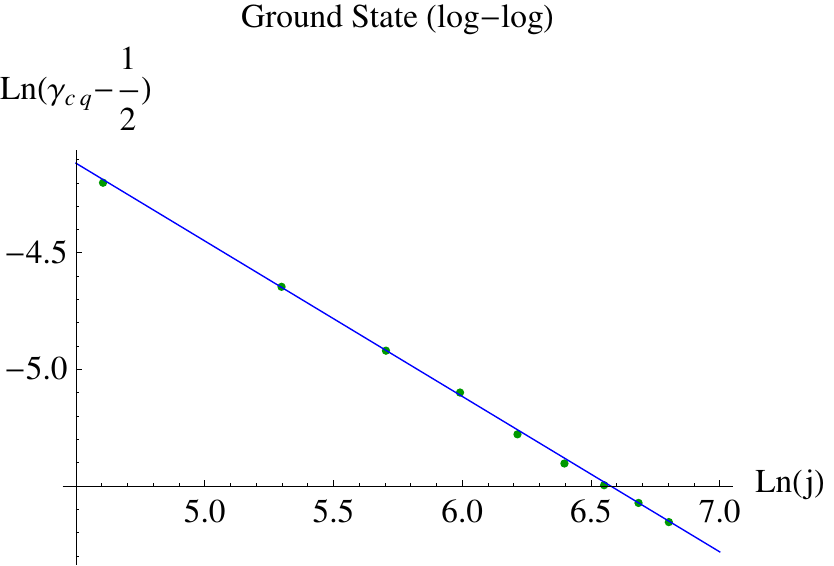}}
	\caption{Logarithmic behaviour of the critical value of the coupling parameter $\gamma_c$ with the number of atoms $N\,(=2j)$, for the quantum ground state. The critical exponent is $-2/3$.}
	\label{gamma_vs_j_loglog}
\end{center}
\end{figure}

The value of $j\,(=N/2)$ was varied from $5$ to $900$ for the ground state, and from $5$ to $450$ for the first excited state. We see that the critical exponent is the same for both curves. A more demanding fit is the logarithmic function of both variables. When we take $j \in [100,\,900]$ for the ground state we get the log-log plot shown in Fig.~\ref{gamma_vs_j_loglog}. Our best fit is
\begin{equation}
	\ln[\gamma_c - \frac{1}{2}] = -1.11596 -\frac{2}{3}\,\ln[j]\ ,
\end{equation}
and we note that $e^{-1.11596} = 0.3276\,\simeq 1/3$.

%Figure 8: q_vs_j quantum log-log

\begin{figure}
	\begin{center}
	\scalebox{0.7}{\includegraphics{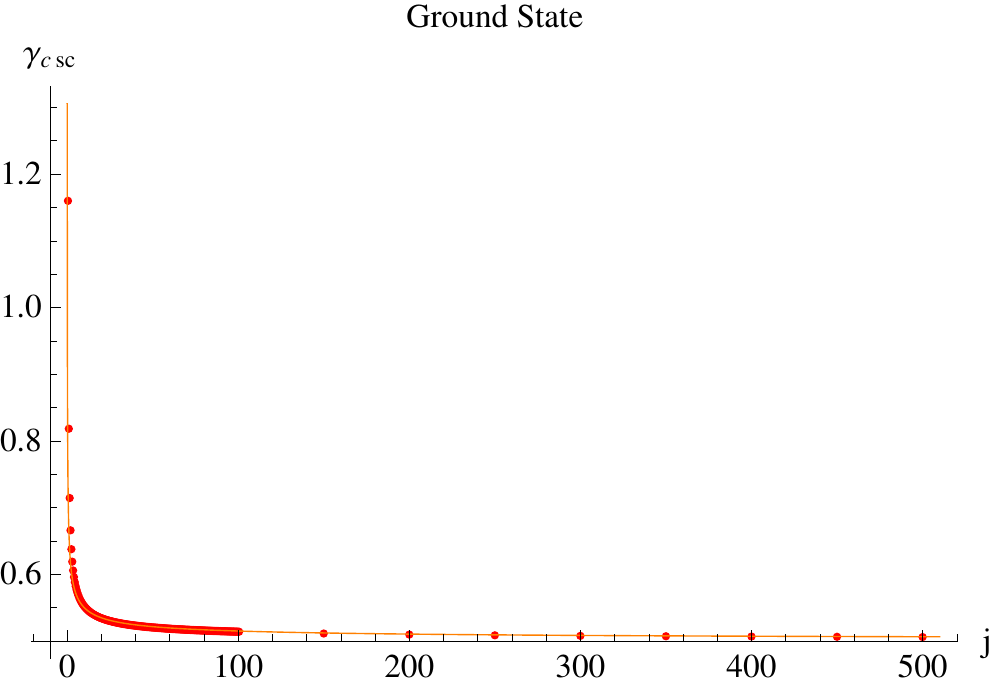}}	 \quad
	\scalebox{0.7}{\includegraphics{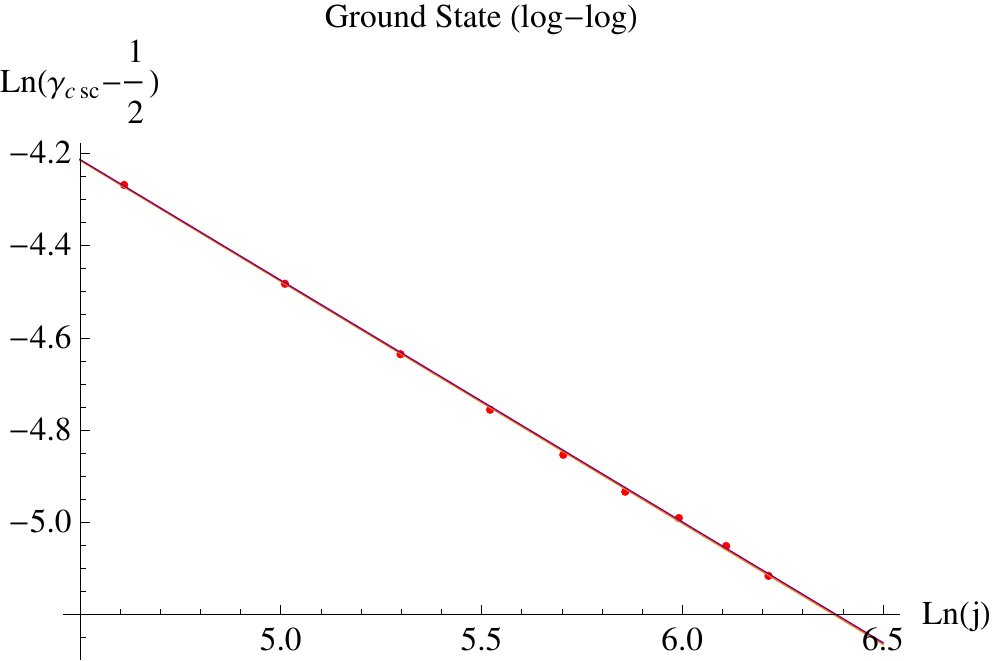}}
	\caption{Left: behaviour of the critical value of the coupling parameter $\gamma_{c\,sc}$ with the number of atoms $N\,(=2j)$, for the SAS ground state. Right: logarithmic behaviour, showing a critical exponent of $-11/21$.}
	\label{gamma_vs_j_sas}
\end{center}
\end{figure}

While the SAS solutions do approximate very well the quantum solutions, except in a small vicinity of the phase transition, when we evaluate the critical coupling parameter coming from SAS as a function of the number of particles, we get a critical exponent of $-11/21$, close but not exactly the appropriate one. In order to obtain this result, recall that the discontinuity in the phase space variables for the even state marks the critical value $\gamma_{c\,sc}$ where the phase transition takes place. We varied $j\,(=N/2)$ from $10$ to $500$ for the ground state. Figure~\ref{gamma_vs_j_sas} (left) shows the results together with the fit (continuous curve)
\begin{equation}
	\gamma_{c\,sc} = \frac{1}{2} + \frac{1}{6}\, j^{-\frac{11}{21}}\ .
\end{equation}
The log-log plot for $j \in [100,\,500]$ shown in Fig.~\ref{gamma_vs_j_sas} (right) confirms this value.  Our best fit is
\begin{equation}
	\ln[\gamma_{c\,sc} - \frac{1}{2}] = \ln[\frac{5}{32}] -\frac{11}{21}\,\ln[j]\ .
\end{equation}
with standard deviation $\sigma = 0.0059$ and confidence interval for the slope (at a confidence level of $0.95$) $m \in [-0.5337, -0.5154]$.

%Figure 9: q_vs_j SAS normal & log-log

\section{Discussion and Conclusions}

We have shown how various mathematical formalisms, specifically the catastrophe formalism and the use of group theory, may aid in the study of non-integrable systems in quantum optics, by comprehensively treating the problem of matter in the presence of a radiation field in a cavity, as modeled by the complete Dicke Hamiltonian.

We approximated the phase transition of the Dicke model for a {\it finite} number $N$ of atoms, via $3$ different methods: i) numerical diagonalisation of the Hamiltonian and the use of the fidelity between neighbouring states; ii) variational (CS) test states that are a direct product of coherent Heisenberg-Weyl ($HW(1)$)-states (for the electromagnetic field), and $SU(2)$-states (for the atomic field); and iii) the use of projection operators on the coherent states in (ii) to obtain states which obey the parity symmetry in the total excitation number, present in the Hamiltonian. We saw that the {\it symmetry-adapted} states (SAS) in (iii) allow for the description of the dependence of observables in the number of particles $N$ for the exact quantum ground and first-excited states.

The three different approaches mentioned above lead to universal parametric curves for the expectation value of the first quadrature of the electromagnetic field (expectation value of the number operator), as a function of the atomic relative population. These are valid for {\it all} values of the matter-field coupling parameter, {\it and} valid for both the ground and first-excited states.

Finally, using these mathematical tools, the critical value of the atom-field coupling parameter $\gamma_{c}$ was found as a function of the number of atoms $N$, from which its critical exponent was derived. This is useful to describe the behaviour of the system near its phase transition.

%%%%%%%%%%%%%%%%%%%%%%%%%%%%%%%%%%%%%%%%%%%%%%%%
%% BACKMATTER
%%%%%%%%%%%%%%%%%%%%%%%%%%%%%%%%%%%%%%%%%%%%%%%%

This work was partially supported by CONACyT-M\'exico (under project 101541), and DGAPA-UNAM (under projects IN102811, IN102109).


\section*{References}
\begin{thebibliography}{99}

\bibitem{power}
Power E A and Zineau S 1959 {\it Phil. Trans. R. Soc.} A {\bf 251} 427

\bibitem{siva}
Sivasubramanian S, Widom A, and Srivastava Y N 2001 {\it Physica} A {\bf 301} 241

\bibitem{dicke}
Dicke R H 1954 {\it Phys. Rev.} {\bf 93} 99

\bibitem{hepp}
Hepp K and Lieb E H 1973 {\it Ann. Phys.} {\bf 76} 360

\bibitem{exp1} Scrinabowitz N, Herman I P, MacGillivray J C and Feld M S 1973 {\it Phys. Rev. Lett.} {\bf 30} 309

\bibitem{wod1} Rzazewski K, Wodkiewicz K and Zakowicz W 1975 {\it Phys. Rev. Lett.} {\bf 35} 432

\bibitem{knight} Knight J M, Aharonov Y and Hsieh G T C 1978 {\it Phys. Rev.} A {\bf 17} 1454; Bialynicki-Birula I and Rzazewski K 1979 {\it Phys. Rev.} A {\bf 19} 301; Gawedzki K and Rzazewski K 1981 {\it Phys. Rev. A} {\bf 23} 2134

\bibitem{domokos}
Vukics A and Domokos P 2012 {\it Adequacy of the Dicke model in cavity QED: 
a counter-``no-go'' statement} arXiv:1206.0752v2 [quant-ph]

\bibitem{baumann}
Baumann K, Guerlin C, Brennecke F and Esslinger T 2010 {\it Nature} {\bf 464} 1301

\bibitem{nagy}
Nagy D, Konya G, Szirmai G and Domokos P 2010 {\it Phys. Rev. Lett.} {\bf104} 130401

\bibitem{canela}
Canela V S C 2012 {\it Haciendo investigaci\'on en \'optica cu\'antica utilizando OpenKet} B.Sc. Thesis, National University of Mexico

\bibitem{tavis}
Tavis M and Cummings F W 1969 {\it Phys. Rev.} {\bf 170} 379; {\bf 188}, 692

\bibitem{scrip}
Casta\~nos O, L\'opez-Pe\~na R, Nahmad-Achar E, Hirsch J G, L\'opez-Moreno E and Vitela J E 2009 {\it Physica Scripta} {\bf 79} 065405

\bibitem{scrip2}
Casta\~nos O, Nahmad-Achar E, L\'opez-Pe\~na R and Hirsch J G 2009 {\it Physica Scripta} {\bf 80} 055401

\bibitem{vaxjo2010}
Nahmad-Achar E, Casta\~nos O, L\'opez-Pe\~na R and Hirsch J G 2012 {\it Quantum Behaviour Mirrored by Semi-Classical States} in Foundations of Probability and Physics - 6 (M. D'Ariano et al., eds.) AIP Conference Proceedings {\bf 1424} 251

\bibitem{dukelsky}
Dukelsky J, Dussel G G, Esebbag C and Pittel S 2004 {\it Phys. Rev. Lett.} {\bf 93} 050403

\bibitem{brennecke}
Brennecke F, Donner T, Ritter S, Bourdel T, Koehl M and Esslinger T 2007 {\it Nature} {\bf 450} 268

\bibitem{jc1963}
Jaynes E T and Cummings F W 1963 {\it Proc. IEEE} {\bf 51} 89

\bibitem{eberly}
Eberly J H, Narozhny N B and Sanchez-Mondragon J J 1980 {\it Phys. Rev. Lett.} {\bf 44} 1323

\bibitem{manko}
{\it Theory of Nonclassical States of Light} 2003 (V.V. Dodonov and V.I. Man'ko, eds.) Taylor \& Francis, London

\bibitem{papercorto}
Nahmad-Achar E, Casta\~nos O, L\'opez-Pe\~na R and Hirsch J G 2011 {\it Phys. Rev.} A {\it Rapid Comm.} {\bf 83} 051601(R)

\bibitem{paperextenso}
Nahmad-Achar E, Casta\~nos O, L\'opez-Pe\~na R and Hirsch J G 2011 {\it Phys. Rev.} A {\bf 84} 013819

\bibitem{leviatan}
Leviatan A 2006 {\it Phys. Rev.} C {\bf 74} 051301(R); Cejnar P 2011 {\it J. Phys. Conf. Ser.} {\bf 322} 012012; Rossignoli R and Plastino A 1985 {\it Phys. Rev.} C {\bf 32} 1040

\bibitem{emary}
Emary C and Brandes T 2003 {\it Phys. Rev.} E {\bf 67} 066203; {\it Phys. Rev. Lett.} {\bf 90} 044101

\bibitem{hecht}
Hecht K T 1987 {\it The Vector Coherent State Method and Its Applications
to Problems of Higher Symmetries} Springer Verlag (Berlin)

\bibitem{gilmore}
Gilmore R 1981 {\it Catastrophe Theory for scientists and engineers} Wiley (New York)

\bibitem{universal}
Nahmad-Achar E, Casta\~nos O, L\'opez-Pe\~na R and Hirsch J G 2012 {\it Phys. Rev.} A {\bf 86} 023814

\bibitem{rumania}
Hirsch J G, Casta\~nos O, Nahmad-Achar E and L\'opez-Pe\~na R 2012 {\it Phase transitions with finite atom number in the Dicke Model} Physica Scripta Topical Issue (to be published)

\bibitem{romera}
Romera E, del Real R and Calixto M 2012 {\it Phys. Rev.} A {\bf 85} 053831

\bibitem{zanardi}
This was found previously and independently by Zanardi P and Paunkovi\'c N 2006 {\it Phys. Rev.} E {\bf 74} 031123

\bibitem{gu}
Gu S-J 2010 {\it Int. J. Mod. Phys.} B {\bf 24} 4371

\bibitem{katzgraber}
Katzgraber H G {\it Phase Transitions: Proseminar in Theoretical Physics}, Institut f\"ur Theoretische Physik (ETH Z\"urich) SS07

\end{thebibliography}
\end{document}